\documentclass[twocolumn,amsmath,amssymb,aps, prc, showpacs, lengthcheck, floatfix]{revtex4}
\usepackage{graphicx}
\usepackage[english]{babel}
\usepackage[colorlinks=true]{hyperref}
\usepackage{multirow}
\usepackage{morefloats}

\graphicspath{{img/}}

\begin{document}

\title{Light-ion production in the interaction of 175 MeV quasi-mono-energetic neutrons with iron and with bismuth}

\author{R. Bevilacqua}\altaffiliation[Present address: ]{European Spallation Source ESS AB, P.O Box 176, SE-221 00 Lund, Sweden}
\author{K. Jansson}
\author{S. Pomp}
\email[Corresponding author. E-mail: ]{stephan.pomp@physics.uu.se}
\author{P. Andersson}
\author{J. Blomgren}
\author{C. Gustavsson}
\author{A. Hjalmarsson}
\author{V. D. Simutkin}
\author{M. \"Osterlund}
\affiliation{Department of Physics and Astronomy, Uppsala University, Box 516, 751 20 Uppsala, Sweden}

\author{A. J. Koning}
\affiliation{Department of Physics and Astronomy, Uppsala University, Box 516, 751 20 Uppsala, Sweden}
\affiliation{Nuclear Research and Consultancy Group NRG, P.O. Box, 1755 ZG Petten, The Netherlands}

\author{A. V. Prokofiev}
\affiliation{Department of Physics and Astronomy, Uppsala University, Box 516, 751 20 Uppsala, Sweden}
\affiliation{The Svedberg Laboratory, Uppsala University, 751 21 Uppsala, Sweden}

\author{M. Hayashi}
\author{S. Hirayama}
\author{Y. Naitou}
\author{Y. Watanabe}
\affiliation{Department of Advanced Energy Engineering Science, Kyushu University, Kasuga, Fukuoka 816-8580, Japan}

\author{U. Tippawan}
\affiliation{Plasma and Beam Physics Research Facility, Department of Physics and Materials Science, Faculty of Science, Chiang Mai University, Chiang Mai 50200, Thailand}

\author{S. G. Mashnik}
\affiliation{Los Alamos National Laboratory, Los Alamos, NM 87545, USA}

\author{L. M. Kerby}
\affiliation{Los Alamos National Laboratory, Los Alamos, NM 87545, USA}
\affiliation{University of Idaho, Moscow, ID 83844, USA}

\author{F.-R. Lecolley}
\author{N. Marie}
\affiliation{LPC, Universit\'e de Caen, 14050 Caen, France
}

\author{J.-C. David}
\author{S. Leray}
\affiliation{CEA-Saclay, Irfu-SPhN, Gif-sur-Yvette, 91191, France}

\date{\today}

\begin{abstract}
Nuclear data for neutron-induced reactions in the intermediate energy range of 20 to 200 MeV are of great importance for the development of nuclear reaction codes since little data exist in that range. Also several different applications benefit from such data, notably accelerator-driven incineration of nuclear waste. The Medley setup was used for a series of measurements of p, d, t, $^3$He and $\alpha$-particle production by 175 MeV quasi-mono-energetic neutrons on various target nuclei. The measurements were performed at the The Svedberg Laboratory in Uppsala, Sweden. Eight detector telescopes placed at angles between 20$^\circ$ and 160$^\circ$ were used. Medley uses the $\Delta E$-$\Delta E$-$E$ technique to discriminate among the particle types and is able to measure double-differential cross sections over a wide range of particle energies. This paper briefly describes the experimental setup, summarizes the data analysis and reports on recent changes in the previously reported preliminary data set on bismuth. Experimental data are compared with INCL4.5-Abla07, MCNP6 using CEM03.03, TALYS and PHITS model calculations as well as with nuclear data evaluations. The models agree fairly well overall but in some cases systematic differences are found.
\end{abstract}

\maketitle
\section{Introduction}
Development of nuclear facilities for accelerator-driven incineration of nuclear waste, particle therapy for cancer treatment, long-range manned space missions, testing and studies of radiation effects in electronic components and systems, and the related dosimetry and radiation safety issues, urge the nuclear data community to provide new and more reliable information on nuclear data for neutron-induced nuclear reactions in the intermediate energy region from 20 to 200\,MeV where experimental data are very scarce. In this energy region all structural properties of the target nucleus may influence the nuclear reaction. Evaluated data libraries need to be verified against experimental data. Neutron-induced reaction data provide benchmark points for theoretical models and help to ensure a good link between low- and high-energy processes, where experimental data and reliable models are available.

Important for the mentioned applications are double-differential cross sections (DDXs) of light-ion production in neutron-induced reactions. To satisfy these needs, a series of measurements has earlier been performed at The Svedberg Laboratory (TSL), Uppsala at energies around 96\,MeV and 175\,MeV \cite{Medley10years}. In this work, DDXs for light ions (p, d, t, $^3$He, and $\alpha$) produced in Fe and Bi by 175-MeV quasi-mono-energetic neutrons were measured using the Medley setup as part of the ANDES project with the special aim to study the amount of tritium production. 

In this paper we summarize the measurement and the analysis procedure and compare the experimental results with model calculations. Preliminary data and comparisons with model calculations have been reported earlier \cite{RiccardoThesis} but the applied thick target correction have since been found to be erroneous in the bismuth case. Here we present the Fe and the revised Bi data. In this paper we compare the experimental data with several model calculations (INCL4.5-Abla07 \cite{incl,abla}, MCNP6 \cite{mcnp-6} using CEM03.03 \cite{mcnp-7,mcnp-8}, TALYS-1.2 \cite{talys} and PHITS \cite{qmd-7,qmd-8}).

\section{Materials and methods} 

\subsection{Experimental setup and data reduction}
We have described the experimental setup and data reduction procedures in a previous work~\cite{MedleyNIMA2011}. Here a brief summary is given.

\begin{figure*}
\includegraphics[width=0.7\textwidth]{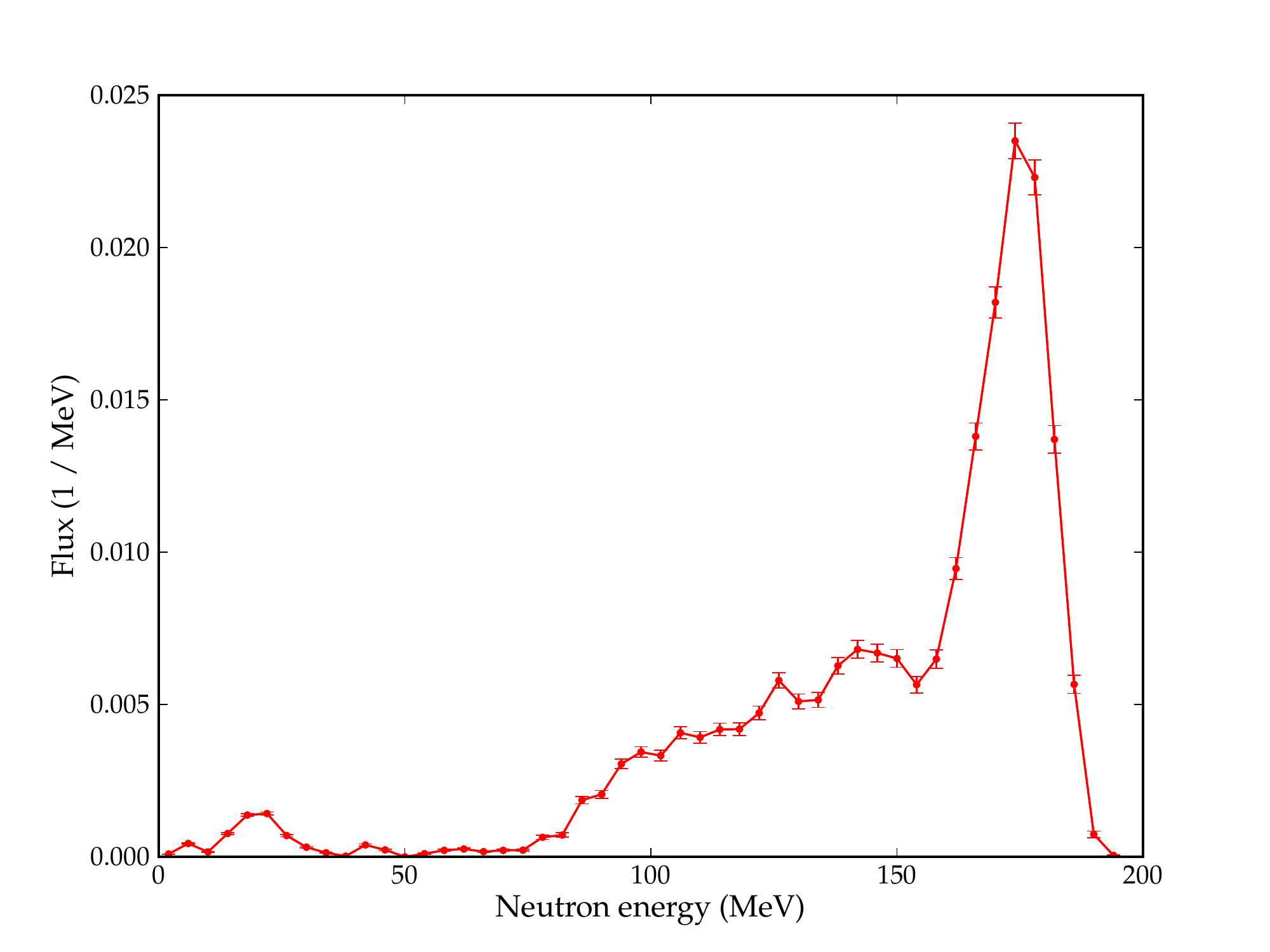}
\caption{\label{fig:acceptedspectrum}The experimental QMN spectrum with the high energy peak at about 175\,MeV.}
\end{figure*}

The experimental work has been conducted at the The Svedberg Laboratory~(TSL) in Uppsala, Sweden, where a quasi-mono-energetic neutron~(QMN) beam is available \cite{TSL}. A~QMN spectrum is characterized by a relatively narrow full-energy peak, followed by a wide low-energy tail. In the present work, full-energy peak neutrons were distributed around 175\,MeV, which is the maximum neutron energy available at~TSL. The~contribution of the low-energy tail was reduced via time-of-flight (TOF) techniques, as described in Ref.~\cite{MedleyNIMA2011}. However, under the present experimental conditions it was not possible to discriminate events induced by 175\,MeV neutrons from events induced by neutrons with energies down to 80 MeV. The experimental QMN spectrum is shown in Fig.~\ref{fig:acceptedspectrum}.

Light-ions were produced in the interaction of the QMN beam with one of the three thin reaction targets, installed inside the Medley scattering chamber; a~0.375\,mm-thick natural-iron target, a~0.5\,mm-thick bismuth target and a 1.0\,mm-thick polyethylene (CH$_2$) target. Charged particles were detected  by eight three-elements telescope detectors, positioned at regular intervals from 20~deg to 160~deg. Each telescope was composed by two fully depleted silicon surface barrier detectors and a CsI(Tl) scintillator.

$\Delta$E-E technique was used for particle identification and total kinetic energy measurement. Normalization to absolute cross section was obtained via the (n,p) cross section (${\sigma}_{\mbox{\tiny{H}}}$), measuring proton recoil from a CH$_2$ target at 20~deg in the laboratory system, in the interaction with 175~MeV peak neutrons. As described in reference~\cite{MedleyNIMA2011}, the absolute double-differential cross section values $\sigma$({\small{E,$\theta$}}) were calculated from experimental data expressed in net counts N({\small{E,$\theta$}}), as
\begin{equation}
	{\sigma}({\small\mbox{E},\theta}) = \frac{\mbox{C} \times \mbox{N}({\small\mbox{E},\theta})}{\Delta\mbox{E}} \frac{{\sigma}{_{\mbox{\tiny{H}}}}({\small \mbox{175 MeV},\mbox{20 deg}})}{\mbox{I}{_{\mbox{\tiny{H}}}}({\small\mbox{175 MeV},\mbox{20 deg}})}
	\label{eq:xs-norm}
	\end{equation}
where ${\sigma}_{\mbox{\tiny{H}}}$=19.9~mb~sr$^{-1}$, I$_{\mbox{\tiny{H}}}$ is the net count of recoil protons from the reference target and $\Delta$E is the energy bin width. The proportionality factor $C$ is determined normalizing the experimental counts to the same neutron fluence, to the number of scattering centres in the targets, and to the detection solid angle.

Instrumental background was measured in a target-out configuration and subtracted from target-in measurements. Nuclear reaction losses in the CsI(Tl) scintillator were corrected by calculating the non-elastic cross sections for each light ion with the respective optical potential, as described in Ref.~\cite{MedleyNIMA2011}. Distortion of the low-energy light-ion production spectra, due to particle and energy losses in the reaction targets, was corrected with the TCORR code~\cite{tcorr} as discussed in Ref.~\cite{RiccardoThesis}. The TCORR code, reimplemented in C++, has been validated by subtracting the energy loss from the corrected spectra and comparing the result to the original measured spectra.

The statistical error in the number of (n,p) counts from the reference CH$_2$ target represents the main source of uncertainties (5\%); other sources are the relative uncertainty from the neutron beam monitors (2~to~3\%) and the uncertainty associated to the reference cross section (2\%). The correction for the thick-target effect contributes to a large systematic uncertainty in low-energy bins (up to 20\%), whereas it does not affect the high-energy end of the measured spectra ($<1\%$).

\subsection{Theoretical models}
All the model calculations were folded with the experimental QMN spectrum depicted in Fig.~\ref{fig:acceptedspectrum} and can therefore be directly compared to the experimental results.

\subsubsection{TALYS calculations}
The TALYS code \cite{talys} integrates the optical model, direct, pre-equilibrium, fission and statistical nuclear reaction models in one calculation scheme and thereby gives a prediction for all the open reaction channels. The purpose of TALYS is to simulate nuclear reactions that involve neutrons, photons, protons, deuterons, tritons, ${}^{3}$He- and $\alpha$-particles, in the 1~keV - 200~MeV energy range. Predicted quantities include integrated, single- and double-differential cross sections, for both the continuum and discrete states, residual production and fission cross sections, gamma-ray production cross sections, etc. For the present work, continuum single- and double-differential cross sections are of interest. To predict these, a calculational scheme is invoked which consists of direct $+$ pre-equilibrium reaction calculation followed by the subsequent compound nucleus decay of all possible residual nuclides by means of the Hauser-Feshbach model.

First, dedicated optical model potentials were developed for both neutrons and protons on ${}^{56}$Fe and ${}^{209}$Bi up to 200 MeV. The used parameters are from the OMP collection of Ref.~\cite{kode}. These potentials provide the necessary reaction cross sections and transmission coefficients for the statistical model calculations. Collective transitions to the continuum were taken into account by contributions from the giant quadrupole and the low-energy and high-energy octupole resonances, whose deformation parameters were determined from the respective energy-weighted sumrules.

For complex particles, the optical potentials were directly derived from the nucleon potentials using Watanabe's folding approach \cite{wata}.

Pre-equilibrium emission takes place after the first stage of the reaction but long before statistical equilibrium of the compound nucleus is attained. It is imagined that the incident particle step-by-step creates more complex states in the compound system and gradually loses its memory of the initial energy and direction. The default pre-equilibrium model of TALYS is the two-component exciton model of Kalbach \cite{kal86}.

The calculated energy spectra were folded with Kalbach's systematics for the angular distributions \cite{kal88} to obtain the double-differential cross sections.

To account for the evaporation peaks in the charged-particle spectra, (which are present, though small), multiple compound emission was treated with the Hauser-Feshbach model. In this scheme, all reaction chains are followed until all emission channels are closed. We adopted Ignatyuk's model \cite{ign75} for the total level density to account for the damping of shell effects at high excitation energies.

For pre-equilibrium reactions involving deuterons, tritons, $^3$He- and $\alpha$-particles, a contribution from the exciton model is automatically calculated with the formalism described above. It is however well-known that for nuclear reactions involving projectiles and ejectiles with different particle numbers, mechanisms like stripping, pick-up and knock-out play an important role and these direct-like reactions are not covered by the exciton model. Therefore, Kalbach \cite{kal05} developed a phenomenological contribution for these mechanisms, which we have included in TALYS. It has been shown, see Table I of Ref. \cite{ker02}, that such methods give a considerable improvement over the older methods (which seemed to result in a consistently strong underestimation of experiment for neutron-induced reactions)

\subsubsection{Modified Quantum Molecular Dynamics}
Quantum Molecular Dynamics (QMD) is a semiclassical simulation method that provides a microscopic description of the evolution of a nucleon many-body system in time \cite{qmd-1,qmd-2,qmd-3}. Each nucleon involved in the reaction is propagated in the nuclear mean field formed by all other nucleons. Stochastic two-body collisions describe the interaction among nucleons. The system evolves for a given time interval, of the order of the nuclear interactions time scale (10$^{-22}$~s); at the end of the simulation, momentum and relative position of each nucleon is verified against conditions defined \textit{a~priori}. Formation of clusters, and emission of single or composite particles are determined based on such conditions. When the compound system reaches equilibrium, the emission of particles is described by an evaporation model. In the present work, the generalized evaporation model (GEM) \cite{qmd-4mcnp-9} is used.

The QMD method has been complemented by a Surface Coalescence Model (SCM) following the observation of a systematic underestimation in the pre-equilibrium production of composite light-ions~\cite{qmd-5,qmd-6}. The SCM describes the formation of clusters in the surface region of the nucleon many-body system: the mechanism hypothesized is that, before this system could reach statistical equilibrium, composite particles may be formed by coalescence to a leading nucleon ready to leave the system itself. Hereafter this calculation model will be referred as MQMD (Modified QMD)-GEM. We have performed MQMD-GEM calculations using a preliminary version \cite{qmd-7} of particle and heavy ion transport code PHITS \cite{qmd-8}.

\subsubsection{MCNP6}
The Monte Carlo particle transport code MCNP6 \cite{mcnp-6} represents a merger of previous MCNP5 and MCNPX codes developed at the Los Alamos National Laboratory. The cascade-exciton model (CEM) is implemented in MCNP6 as its default event generator for nuclear reactions induced by nucleons, pions and $\gamma$ at incident energies below several GeVs. Present calculations were performed with the default event generator CEM03.03 \cite{mcnp-7,mcnp-8}.

The improved cascade-exciton model (CEM) as implemented in the code CEM03.03 \cite{mcnp-7, mcnp-8} assumes that the reactions occur generally in three stages: The first stage is the INC, in which primary particles can be re-scattered and produce secondary particles several times prior to absorption by (or escape from) the nucleus. When the cascade stage of a reaction is completed, CEM03.03 uses the coalescence model to ``create'' high-energy $d$, $t$, $^3$He, and $^4$He particles by final-state interactions among emitted cascade nucleons. The emission of the cascade particles determines the particle-hole configuration, $Z$, $A$, and the excitation energy that is the starting point for the second, pre-equilibrium stage of the reaction. The subsequent relaxation of the nuclear excitation is treated with an improved version of the modified exciton model of pre-equilibrium decay followed by the equilibrium evaporation/fission stage (also called the compound nucleus stage), which is described with an extension of the Generalized Evaporation Models (GEM) code, GEM2, by Furihata \cite{qmd-4mcnp-9}. Generally, all components may contribute to experimentally measurable particle emission spectra and affect the final residual nuclei. But if the residual nuclei after the INC have atomic numbers in the range $A < 13$, CEM03.03 uses the Fermi breakup model \cite{mcnp-10} to calculate their further disintegration instead of using the pre-equilibrium and evaporation models. Fermi breakup is faster to calculate than GEM and gives results similar to the continuation of the more detailed models for much lighter nuclei.

\subsubsection{INCL-ABLA}
INCL4.5-Abla07 is a model combination known as one of the best spallation models on the market regarding nucleon-induced reactions \cite{incl1,incl2}. INCL4.5 is an intranuclear cascade model simulating the first fast phase of the reaction by binary collisions. Once the projectile enters the target nucleus made of nucleons in a Fermi gas, all particles move freely in a nuclear potential. Nucleons, pions and light  charged particles (up to $\alpha$-particles) can be used as projectile.

The detail of the ingredients can be read in \cite{incl}, however two features can be mentioned. The criterion to stop the cascade is based on an auto-coherent stopping time, defined once for all in a previous study \cite{incl4.2}. So no pre-equilibrium step is needed in INCL4.5, since the remnant nucleus is thermally equilibrated. The second point is the surface coalescence model implemented in INCL4.5 leading to the emission of light nuclei ($A\leq 8$). This model is based on the scanning of the phase-space surrounding a nucleon escaping from the nucleus to determine if this nucleon can aggregate some other nucleons.

The statistical deexcitation model Abla07 includes various channels to release the remnant energy. Those that are in competition in the reactions discussed here are mainly the evaporations of light particles from neutron up to $\alpha$-particles. Nevertheless a phenomenological process of multifragmentation is included as well as a fission model. More details can be found in \cite{abla}.

\section{Results and discussion}  
\subsection{Target energy loss correction validation}
\begin{figure*}
\includegraphics[width=0.7\textwidth]{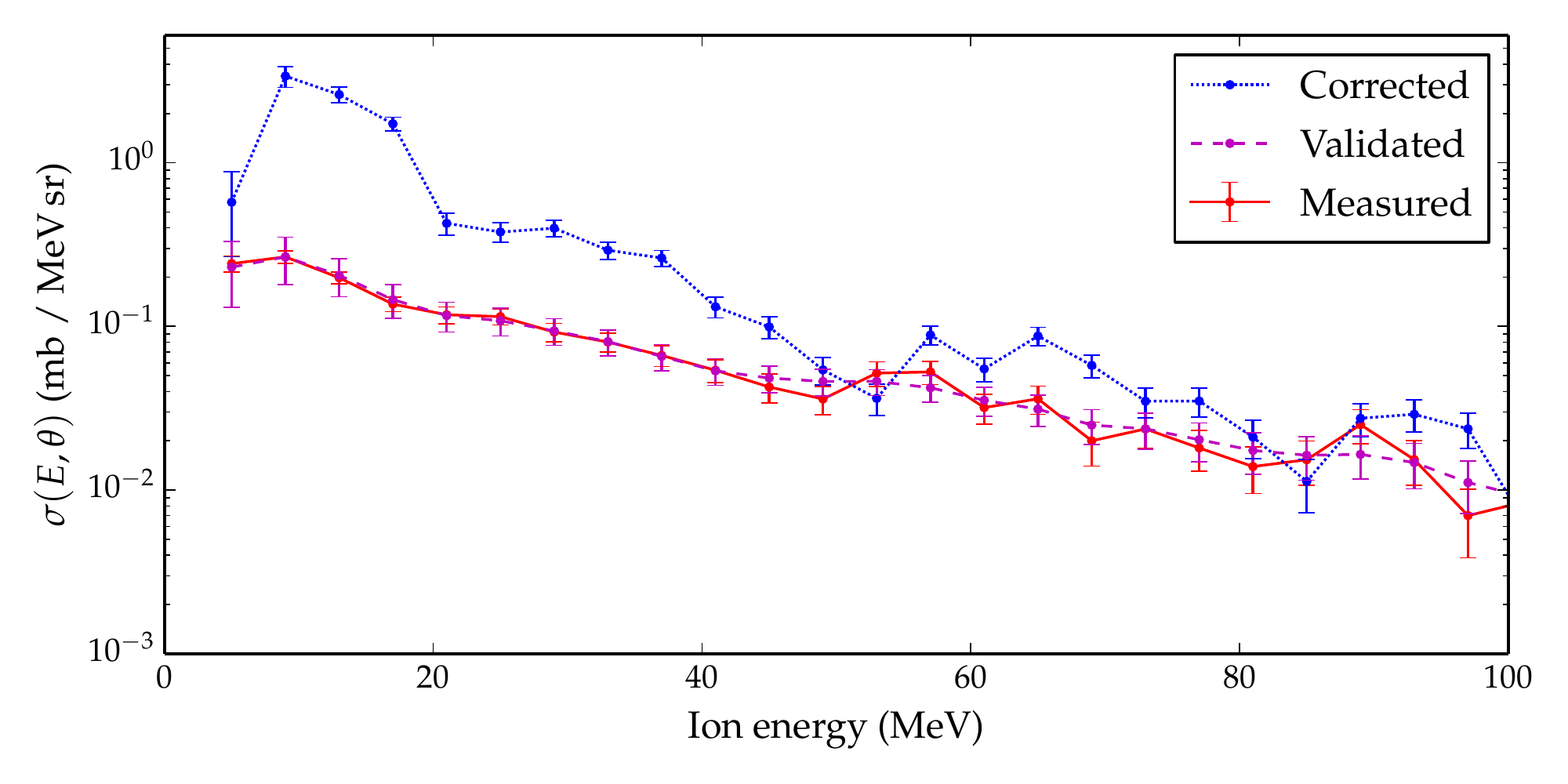}
\caption{\label{fig:fe_validation}The measured double-differential cross section for the Fe(n,x$\alpha$) reaction at 20 degrees, was corrected using the iterative TCORR\cite{tcorr} routine. The energy loss for the corrected double-differential cross section was then calculated and the resulting cross section can be seen to agree well with the original measured data.}
\end{figure*}

The validation was done by taking the TCORR \cite{tcorr} corrected spectrum and simulating the target effects. We could confirm that the TCORR procedure was working since the resulting spectrum turned out to be very similar to the originally measured one. Depicted in Fig.~\ref{fig:fe_validation} is the $\alpha$-particle correction which is the most difficult case since they are most affected by energy loss in the target.

\subsection{Energy cutoffs}
The cutoffs at low energies for the different particles are found in table~\ref{tab:cutoffs}. The low cutoff for $^4$He production on Fe is possible due to the much lower cross section of $^3$He which makes this contribution negligible. One can estimate the non-measured cross section in the cutoff region by using model data in the gap between the production threshold and the cutoff energy. Such estimates have been done for the particle production cross sections. The estimates based on INCL-ABLA, which agrees well with the data overall, are found in table~\ref{tab:cutoffs}. For Fe, the relative magnitude of these corrections is 1~or~2\% while for Bi they are negligible except for the $\alpha$-production cross section correction which still only amounts to 0.4\%.

\begin{table}
\caption{\label{tab:cutoffs}Low energy cutoffs and corresponding corrections to the cross sections. The cross section corrections were calculated based on the INCL-ABLA data below the cutoff energy.}
\begin{tabular}{c|cc|ll}
\toprule
\multirow{2}{*}{Ion} & \multicolumn{2}{c|}{Cutoff energy (MeV)} & \multicolumn{2}{c}{Correction (mbarn)}\\
 & \textbf{Fe} & \textbf{Bi} & \textbf{Fe} & \textbf{Bi} \\
\colrule
p      & 2 & 2  & 15   & 0.056\\
d      & 2 & 3  & 0.24 & 0.0041\\
t      & 2 & 3  & 0.18 & 0.0031\\
$^3$He & 8 & 8  & 0.40 & 0.00 \\
$^4$He & 3 & 14 & 0.15 & 0.36\\
\botrule
\end{tabular}
\end{table}

\subsection{Comparisons to models}
\subsubsection{Double-differential cross sections}
\begin{figure*}[htbp]
\includegraphics[width=0.7\textwidth]{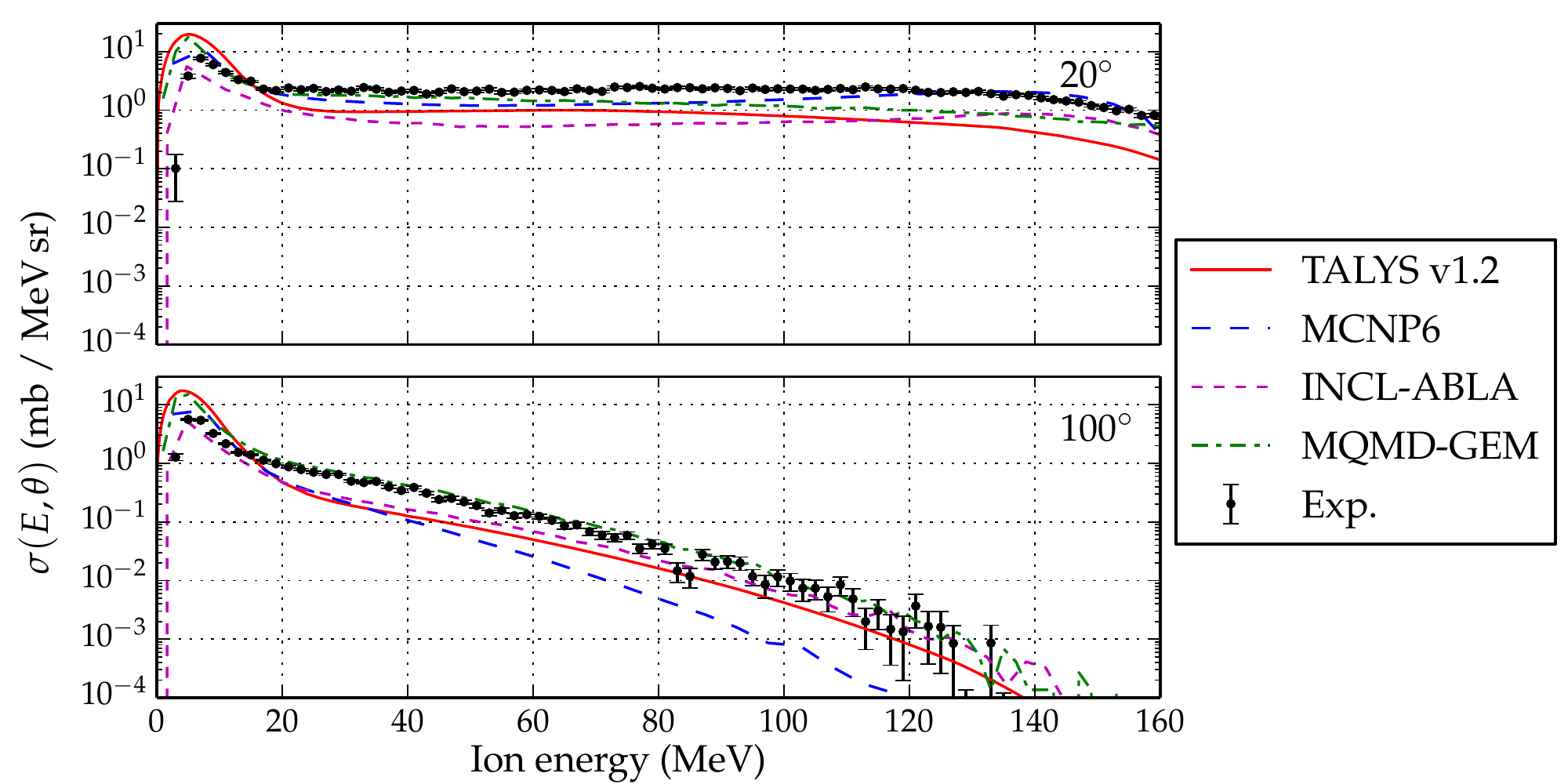}
\caption{\label{fig:fe_p_ddxs}Double differential cross sections for the Fe(n,xp) reaction at the laboratory angles 20$^\circ$ and 100$^\circ$ compared to various models.}
\end{figure*}

\begin{figure*}[htbp]
\includegraphics[width=0.7\textwidth]{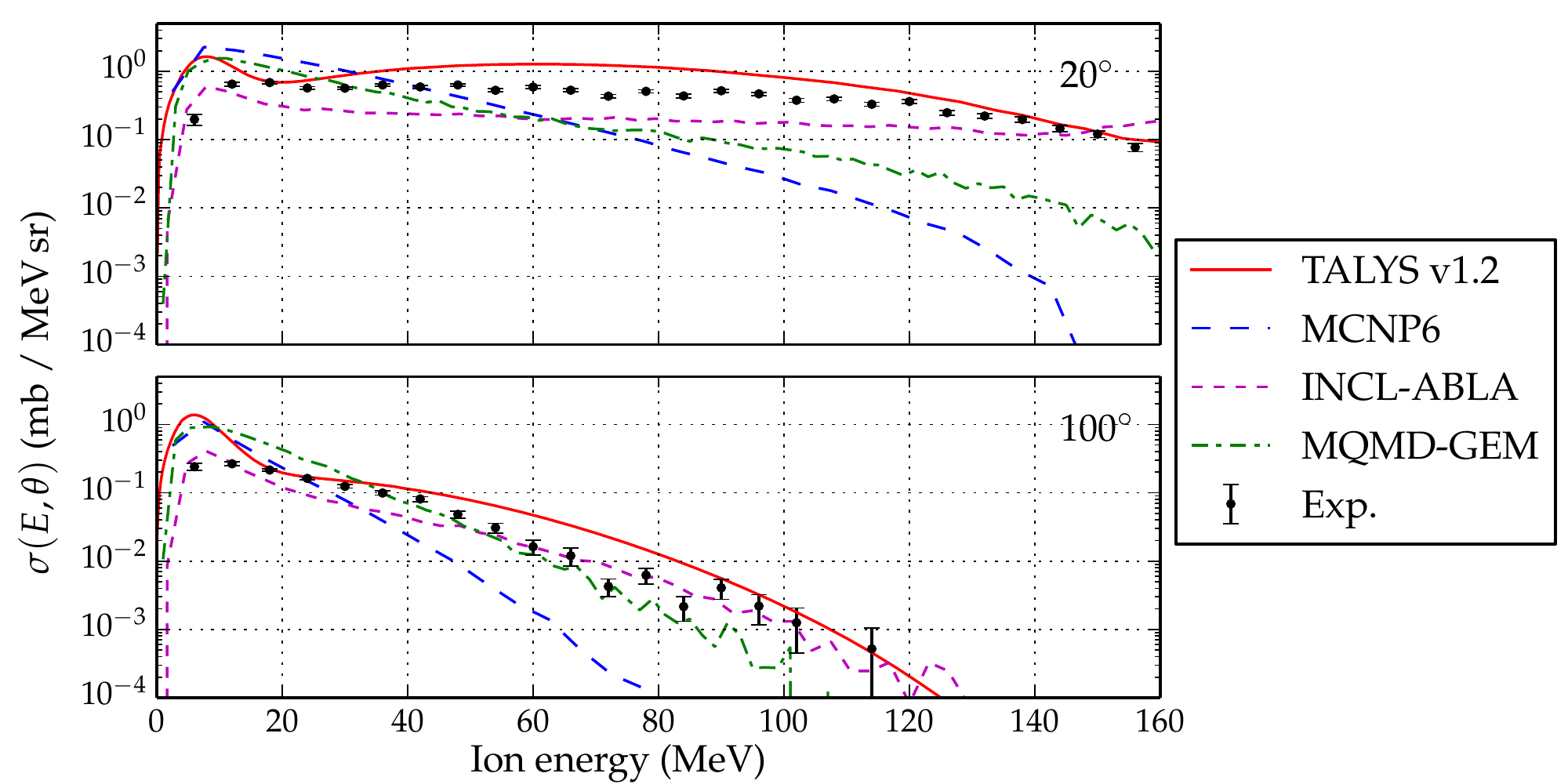}
\caption{\label{fig:fe_d_ddxs}Double differential cross sections for the Fe(n,xd) reaction at the laboratory angles 20$^\circ$ and 100$^\circ$ compared to various models.}
\end{figure*}

\begin{figure*}[htbp]
\includegraphics[width=0.7\textwidth]{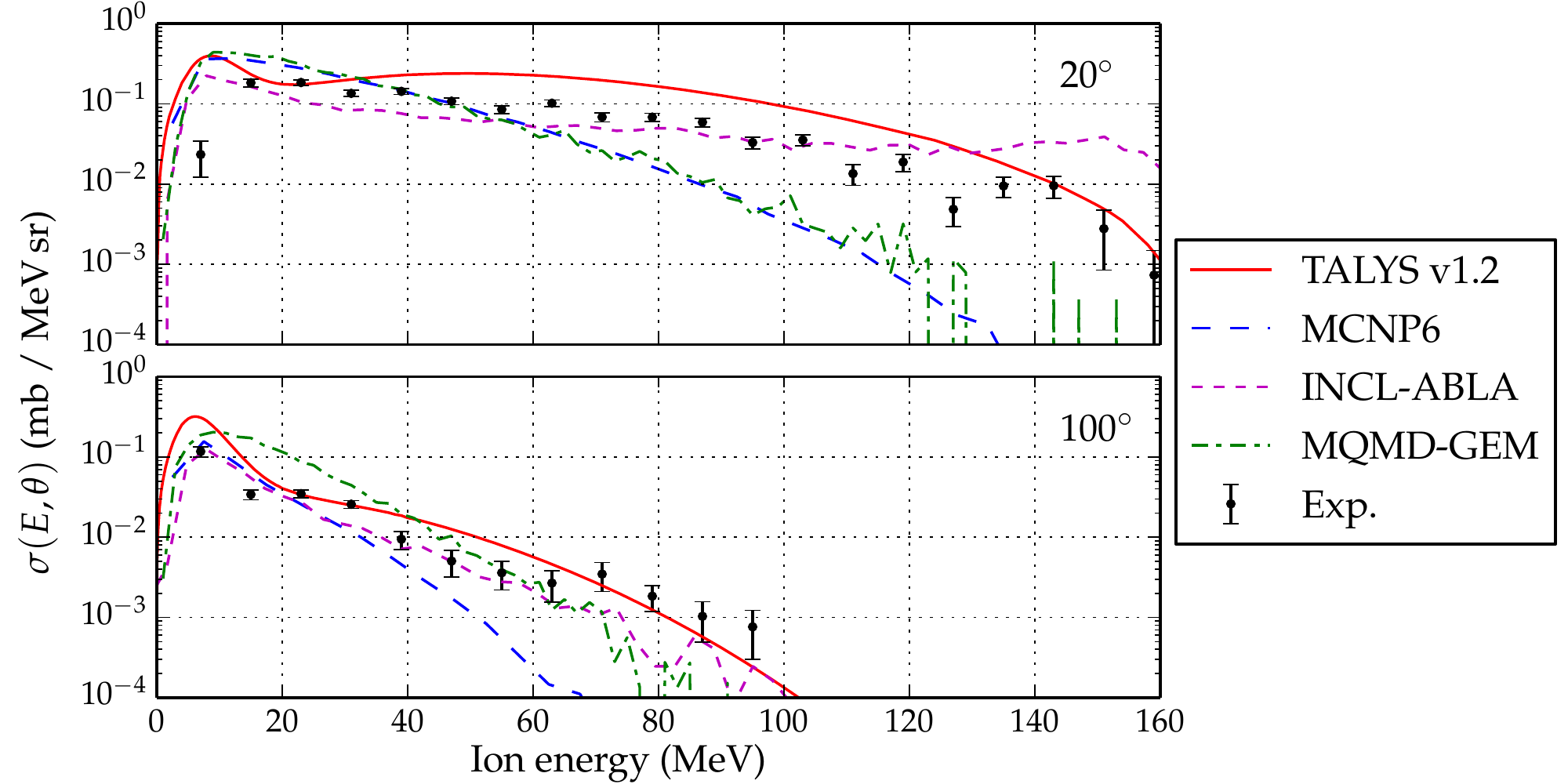}
\caption{\label{fig:fe_t_ddxs}Double differential cross sections for the Fe(n,xt) reaction at the laboratory angles 20$^\circ$ and 100$^\circ$ compared to various models.}
\end{figure*}

\begin{figure*}[htbp]
\includegraphics[width=0.7\textwidth]{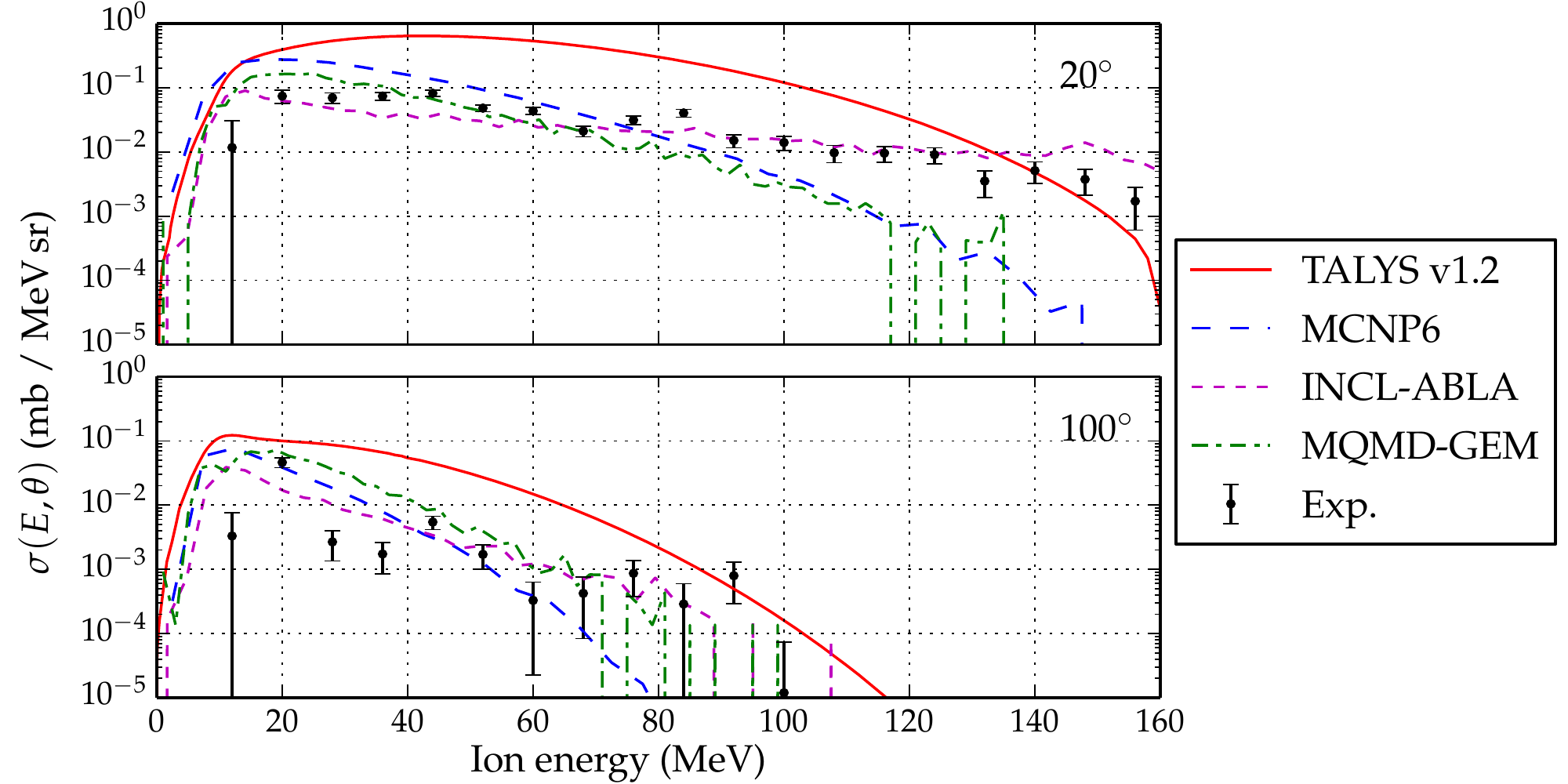}
\caption{\label{fig:fe_3he_ddxs}Double differential cross sections for the Fe(n,x$^3$He) reaction at the laboratory angles 20$^\circ$ and 100$^\circ$ compared to various models.}
\end{figure*}

\begin{figure*}[htbp]
\includegraphics[width=0.7\textwidth]{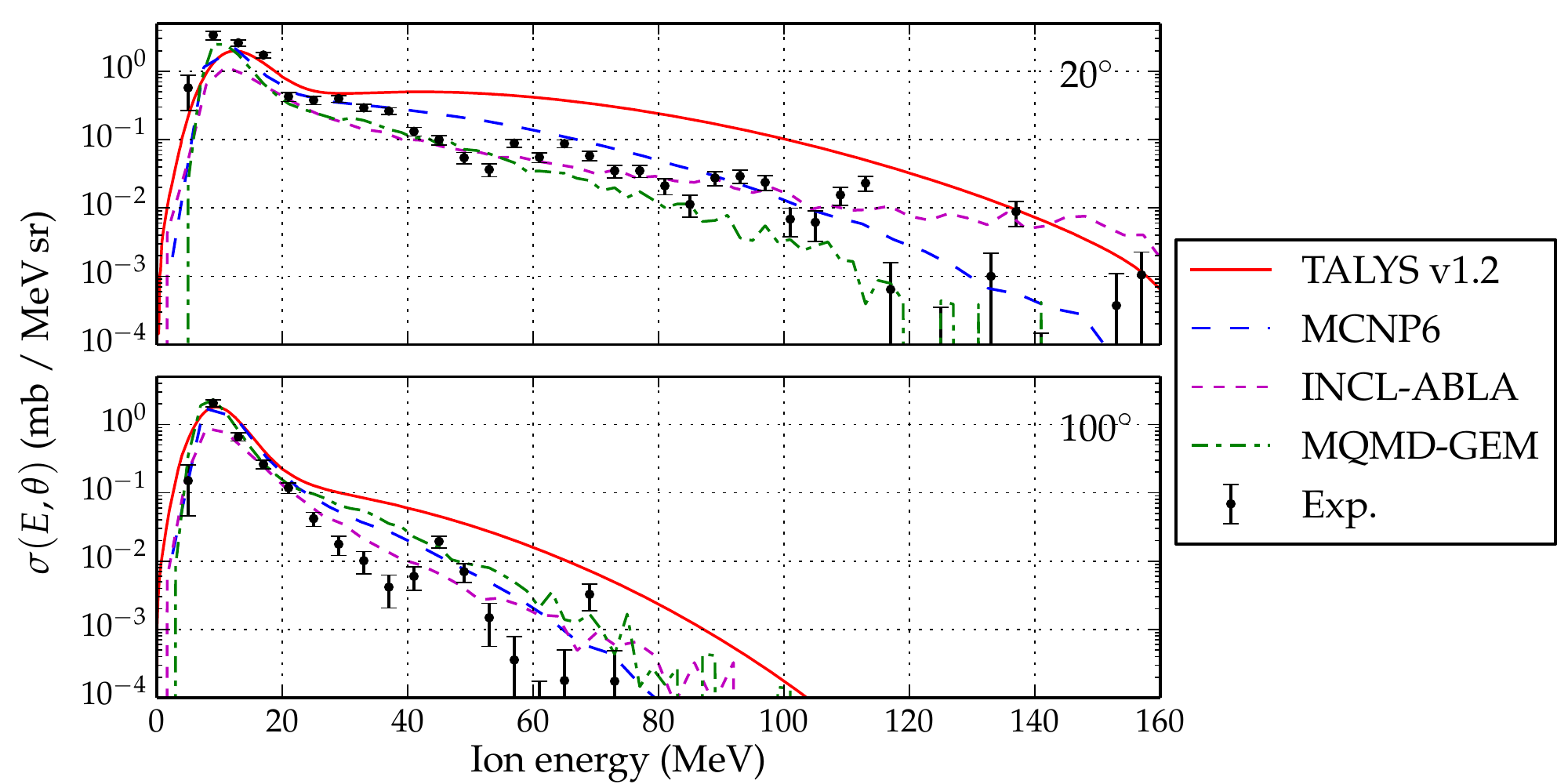}
\caption{\label{fig:fe_4he_ddxs}Double differential cross sections for the Fe(n,x$\alpha$) reaction at the laboratory angles 20$^\circ$ and 100$^\circ$ compared to various models.}
\end{figure*}

\begin{figure*}[htbp]
\includegraphics[width=0.7\textwidth]{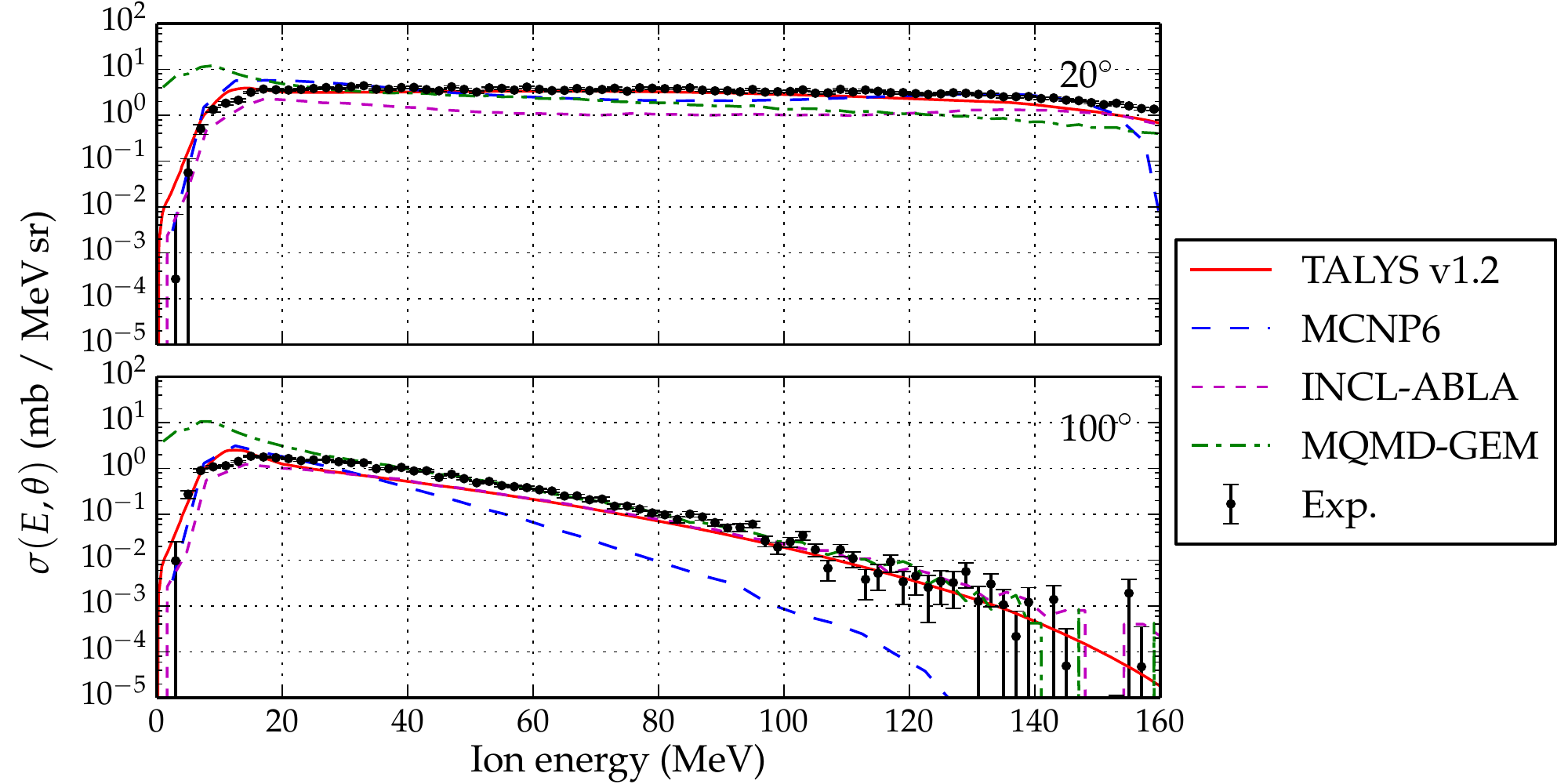}
\caption{\label{fig:bi_p_ddxs}Double differential cross sections for the Bi(n,xp) reaction at the laboratory angles 20$^\circ$ and 100$^\circ$ compared to various models.}
\end{figure*}

\begin{figure*}[htbp]
\includegraphics[width=0.7\textwidth]{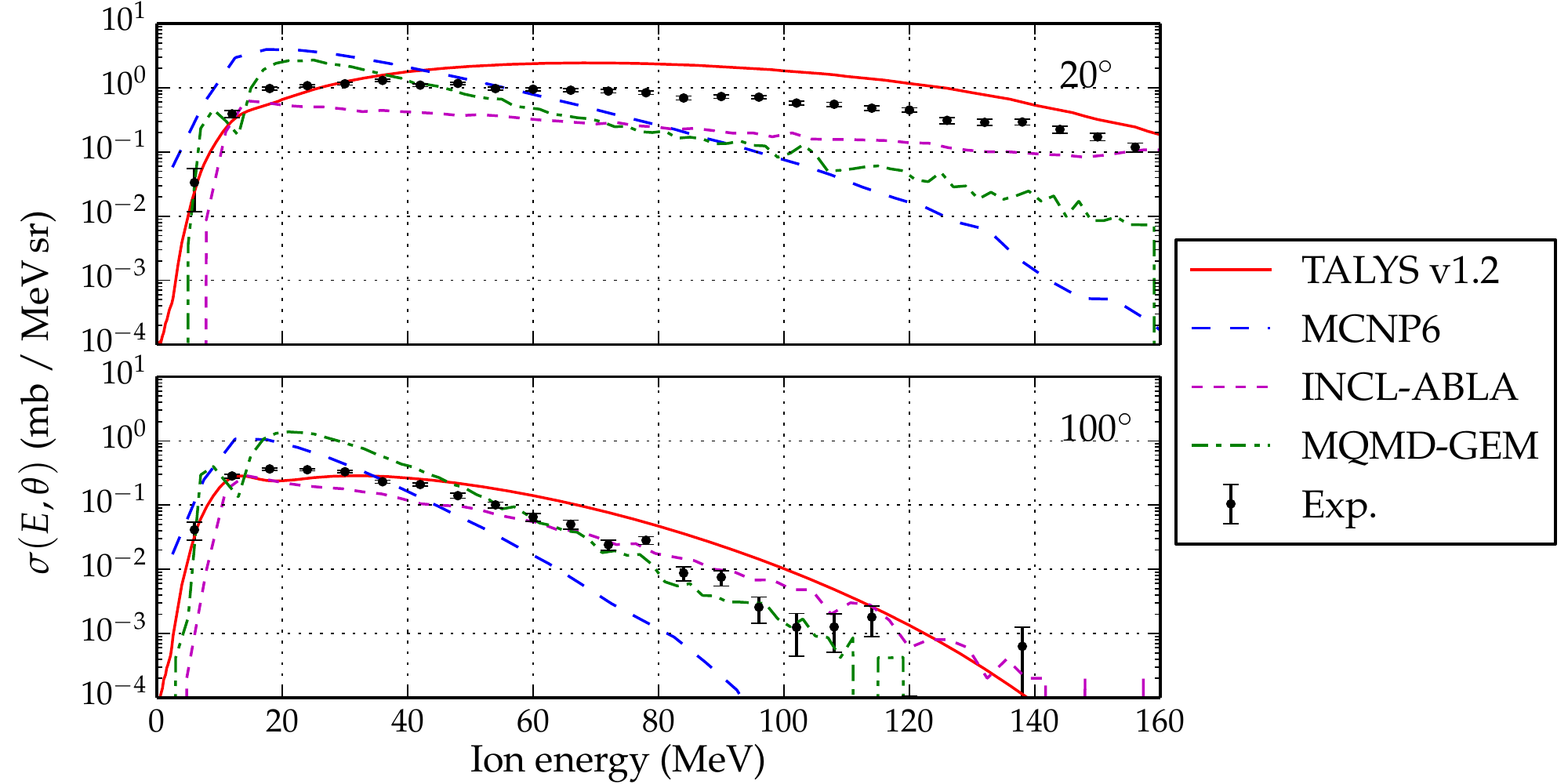}
\caption{\label{fig:bi_d_ddxs}Double differential cross sections for the Bi(n,xd) reaction at the laboratory angles 20$^\circ$ and 100$^\circ$ compared to various models.}
\end{figure*}

\begin{figure*}[htbp]
\includegraphics[width=0.7\textwidth]{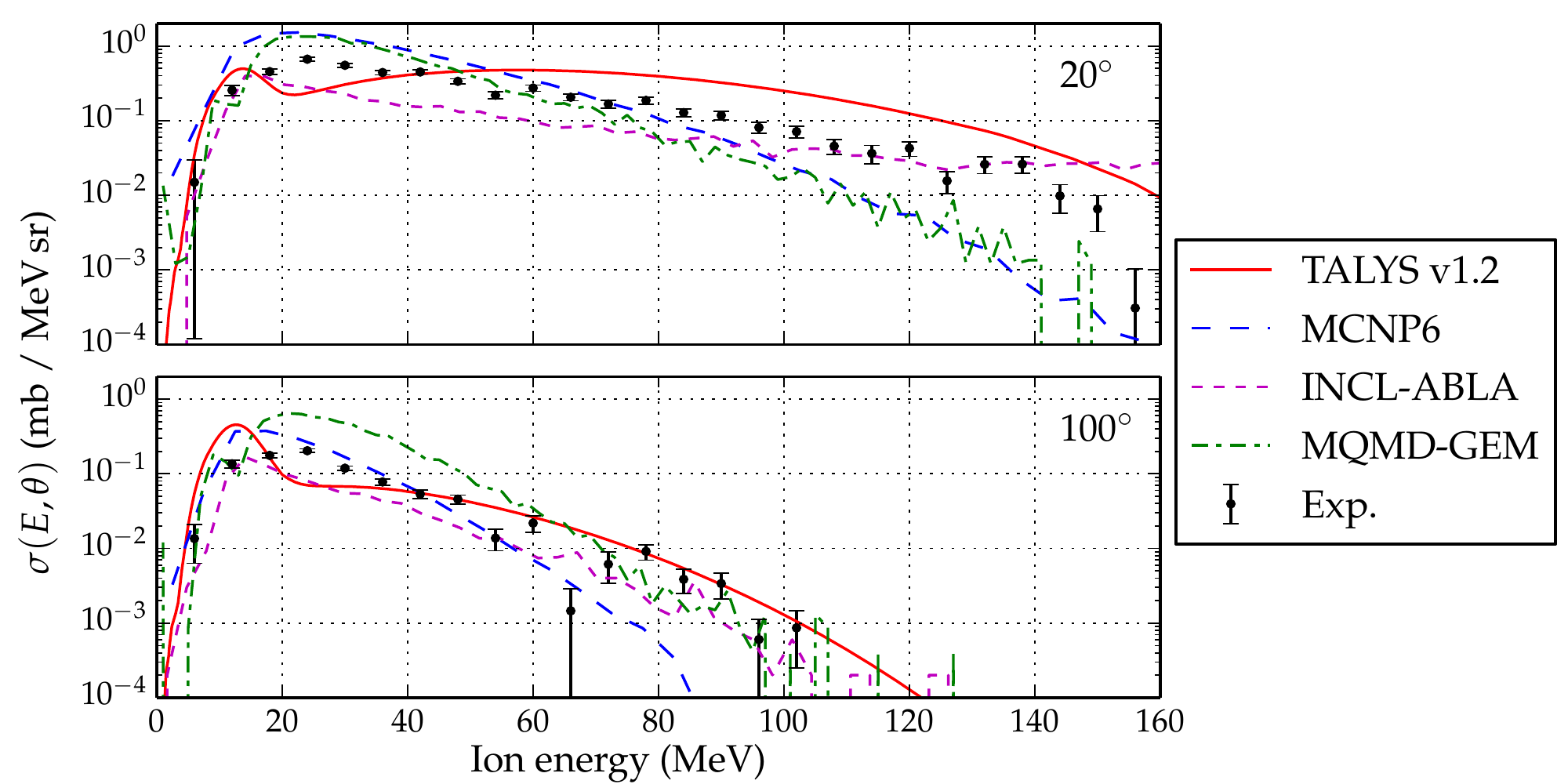}
\caption{\label{fig:bi_t_ddxs}Double differential cross sections for the Bi(n,xt) reaction at the laboratory angles 20$^\circ$ and 100$^\circ$ compared to various models.}
\end{figure*}

\begin{figure*}[htbp]
\includegraphics[width=0.7\textwidth]{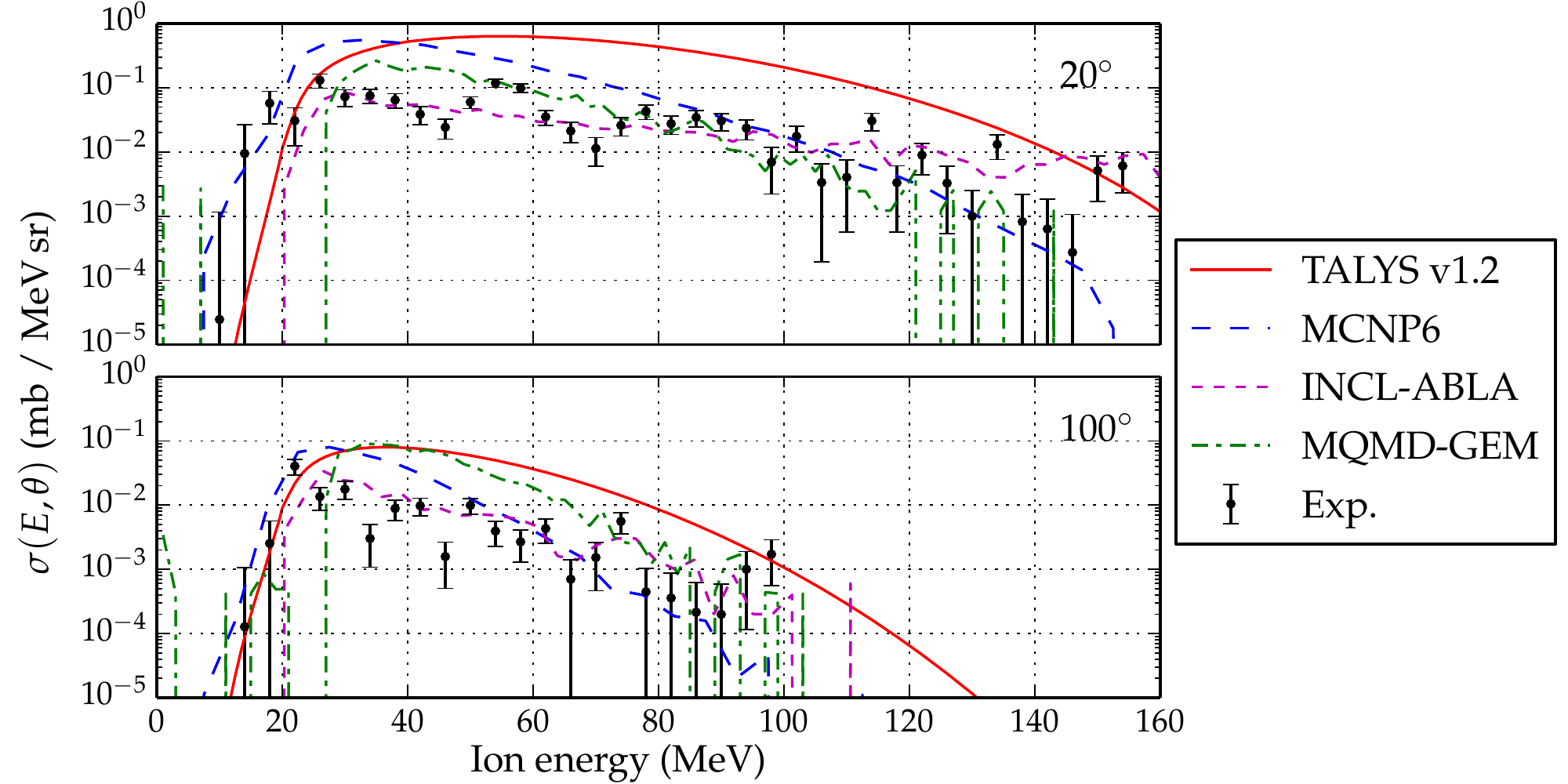}
\caption{\label{fig:bi_3he_ddxs}Double differential cross sections for the Bi(n,x$^3$He) reaction at the laboratory angles 20$^\circ$ and 100$^\circ$ compared to various models.}
\end{figure*}

\begin{figure*}[htbp]
\includegraphics[width=0.7\textwidth]{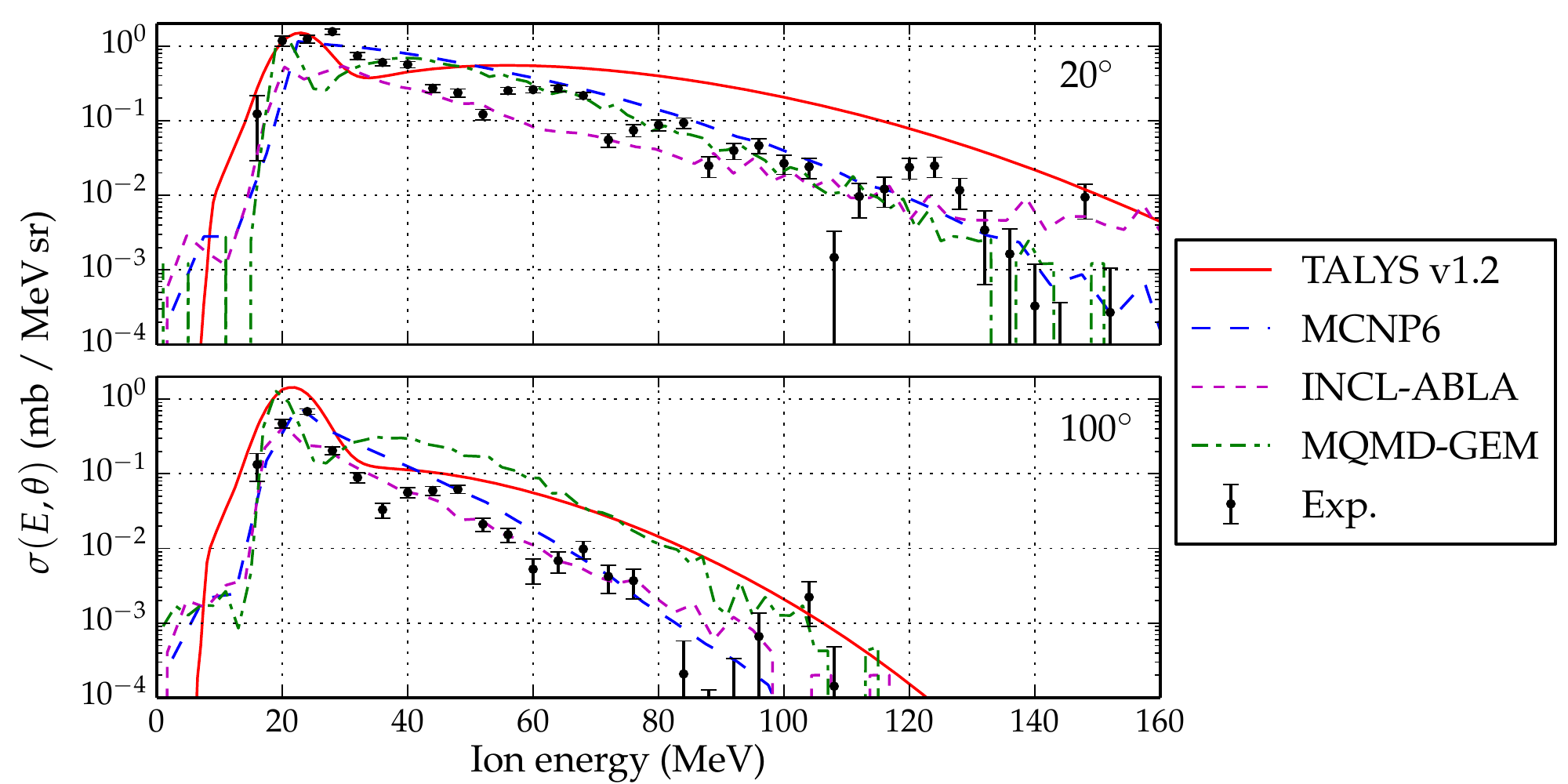}
\caption{\label{fig:bi_4he_ddxs}Double differential cross sections for the Bi(n,x$\alpha$) reaction at the laboratory angles 20$^\circ$ and 100$^\circ$ compared to various models.}
\end{figure*}

Figures~\ref{fig:fe_p_ddxs}-\ref{fig:bi_4he_ddxs} show the experimental results for the double-differential cross sections at 20$^\circ$ and 100$^\circ$ on iron and bismuth respectively. 

\begin{figure*}[htbp]
\includegraphics[width=0.7\textwidth]{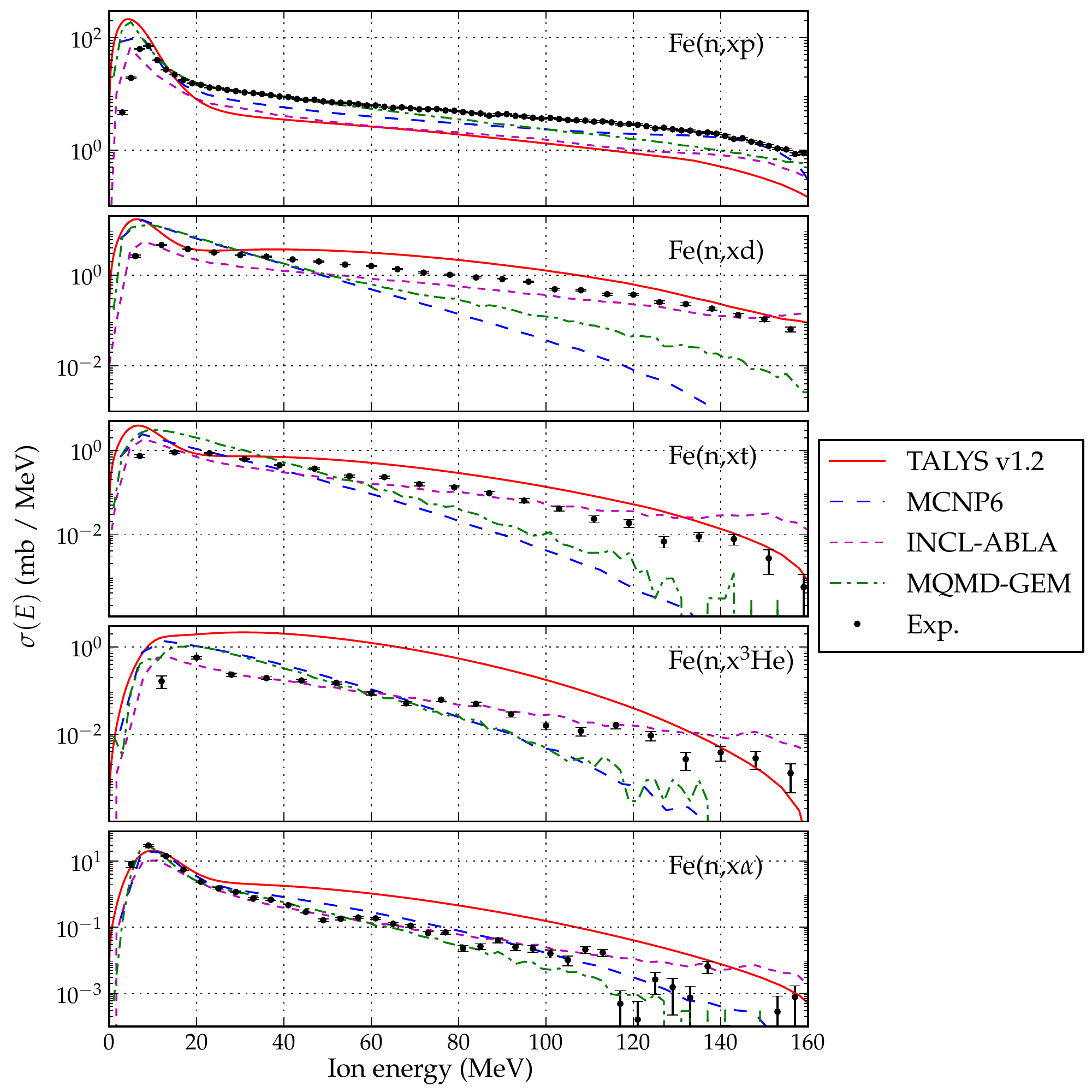}
\caption{\label{fig:fe_edxs}Energy differential cross sections for the Fe(n,xp) reaction compared to various models.}
\end{figure*}

\begin{figure*}[htbp]
\includegraphics[width=0.7\textwidth]{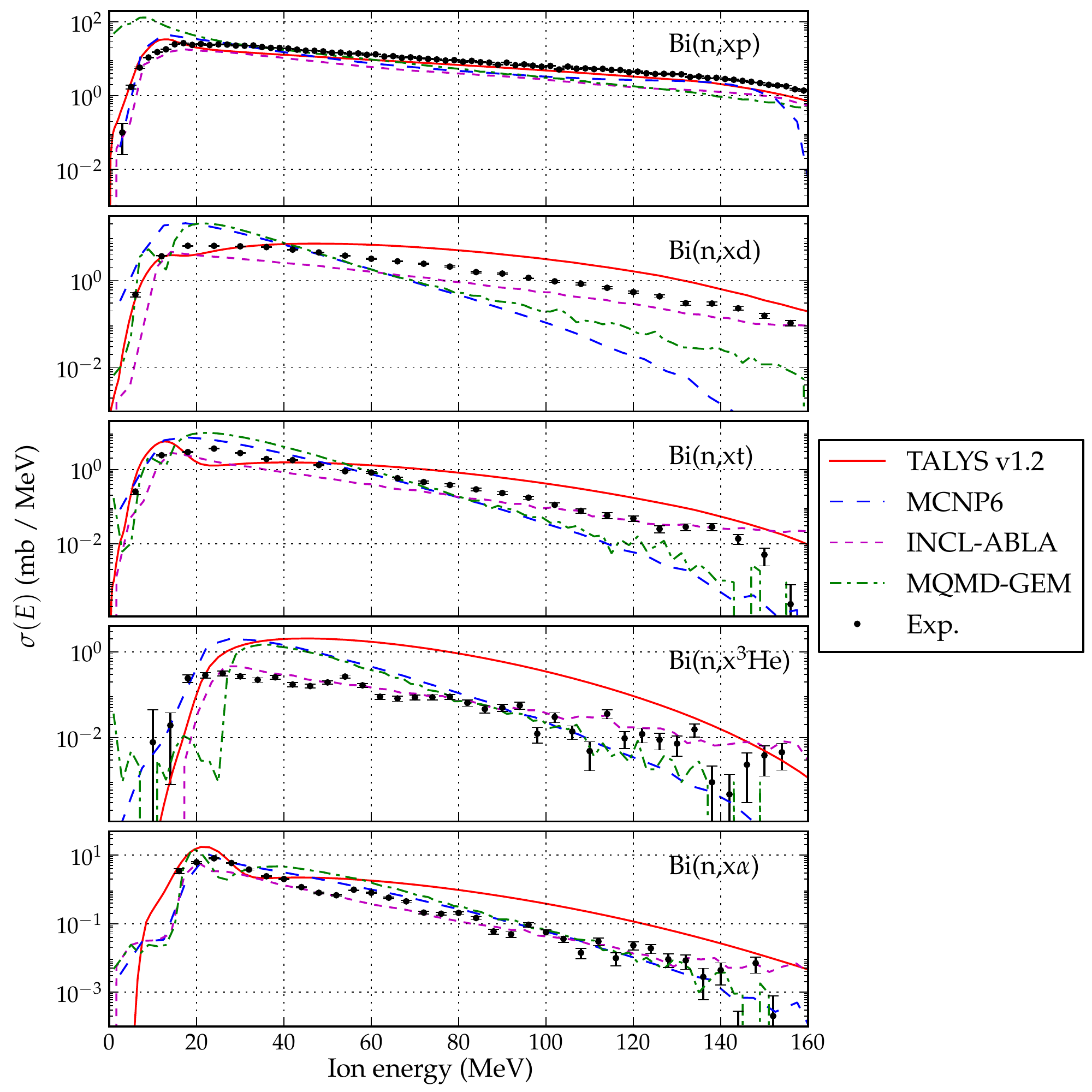}
\caption{\label{fig:bi_edxs}Energy differential cross sections for the Bi(n,xp) reaction compared to various models.}
\end{figure*}

Figures~\ref{fig:fe_p_ddxs} and~\ref{fig:bi_p_ddxs} show a comparison between the double-differential (n,xp) experimental spectra and the calculations from the various models. The Fe(n,xp) spectra are underestimated in the high energy range but overestimated for low particle energies. It is noted that the TALYS-1.2 calculations give a good description of the Bi(n,xp) spectra. The deuteron production calculations lie above the experimental data in both the Fe and Bi cases except for the low energy range in Bi(n,xd), see Figs.~\ref{fig:fe_d_ddxs} and~\ref{fig:bi_d_ddxs}. For triton, $^3$He and $\alpha$-ejectiles, TALYS calculations overestimate the cross section compared to the experimental data for both Fe (Figs.~\ref{fig:fe_t_ddxs}-\ref{fig:fe_4he_ddxs}) and Bi (Figs.~\ref{fig:bi_t_ddxs}-\ref{fig:bi_4he_ddxs}). The largest differences to the data are found for the $^3$He production, while the production of low energy $\alpha$-particles agrees well with the experimental data. The TALYS trends for double-differential cross sections are also visible for the angle-integrated spectra, see Figs.~\ref{fig:fe_edxs} and~\ref{fig:bi_edxs}.

\begin{figure*}[htbp]
\includegraphics[width=0.7\textwidth]{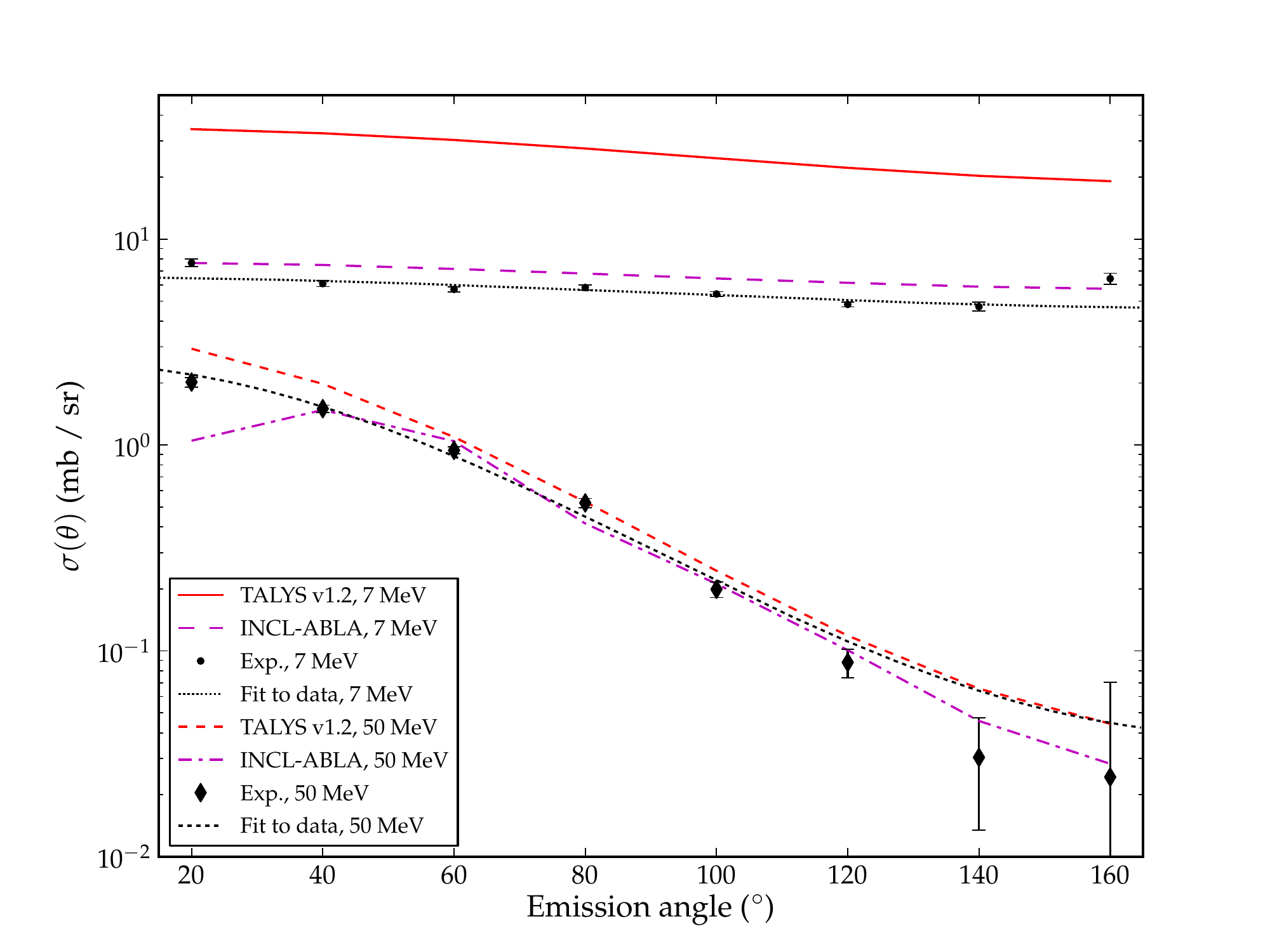}
\caption{\label{fig:fe_p_adxs1}Angular differential cross sections for the Fe(n,xp) reaction at the secondary particle energies 7\,MeV and 50\,MeV compared to TALYS and INCL-ABLA.}
\end{figure*}

\begin{figure*}[htbp]
\includegraphics[width=0.7\textwidth]{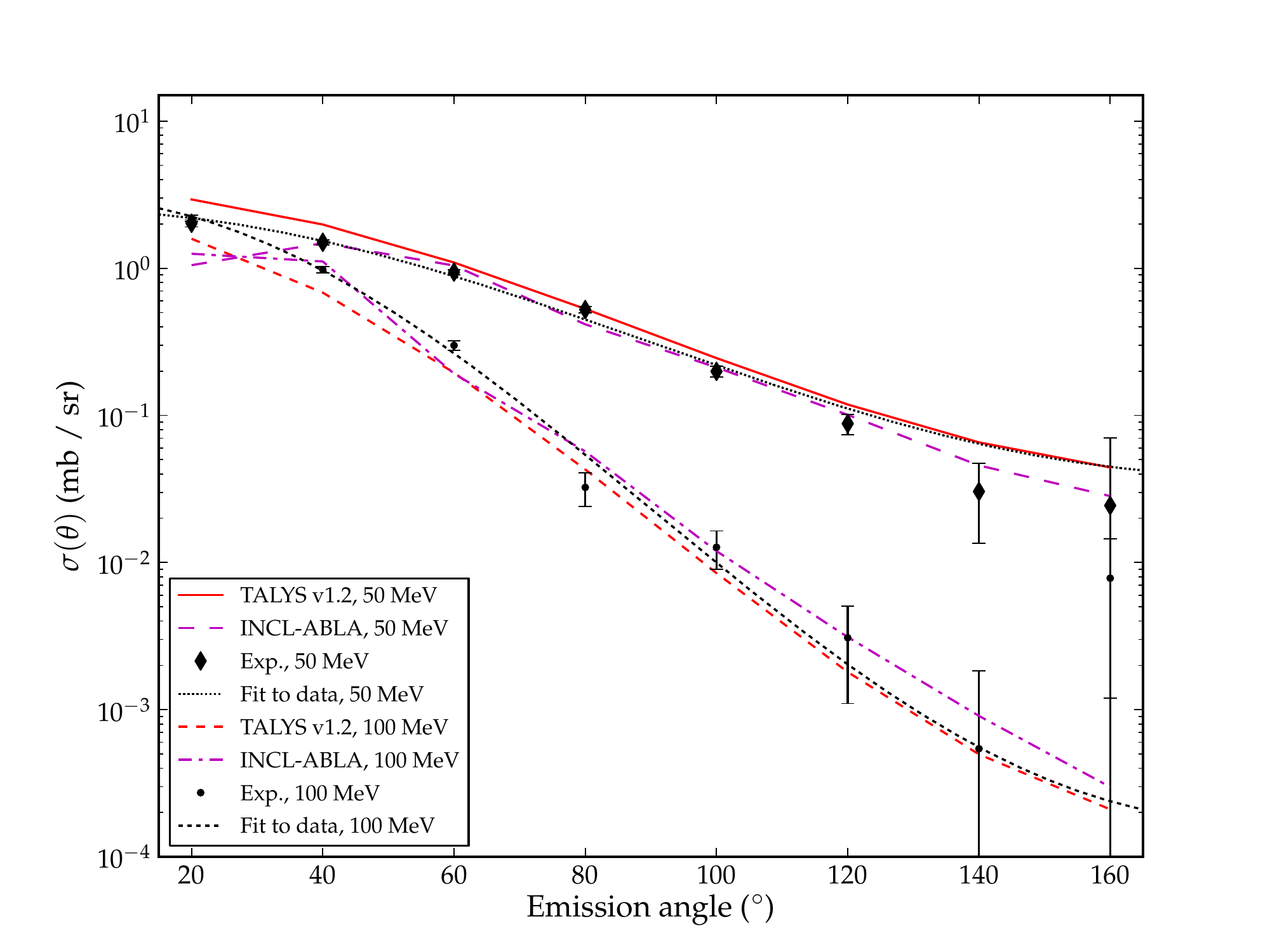}
\caption{\label{fig:fe_p_adxs2}Angular differential cross sections for the Fe(n,xp) reaction at the secondary particle energies 50\,MeV and 100\,MeV compared to TALYS and INCL-ABLA.}
\end{figure*}

\begin{figure*}[htbp]
\includegraphics[width=0.7\textwidth]{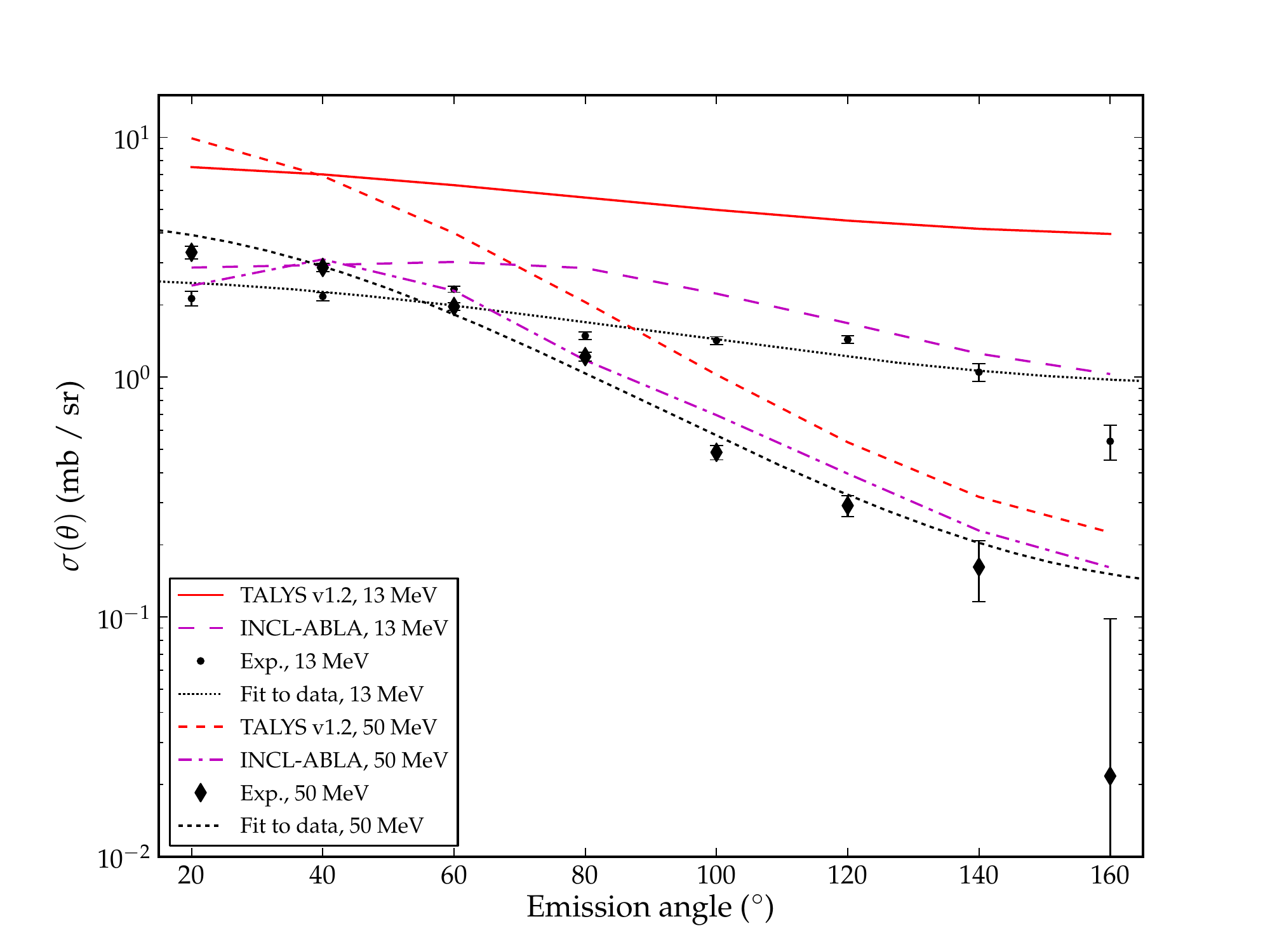}
\caption{\label{fig:bi_p_adxs1}Angular differential cross sections for the Bi(n,xp) reaction at the secondary particle energies 13\,MeV and 50\,MeV compared to TALYS and INCL-ABLA.}
\end{figure*}

\begin{figure*}[htbp]
\includegraphics[width=0.7\textwidth]{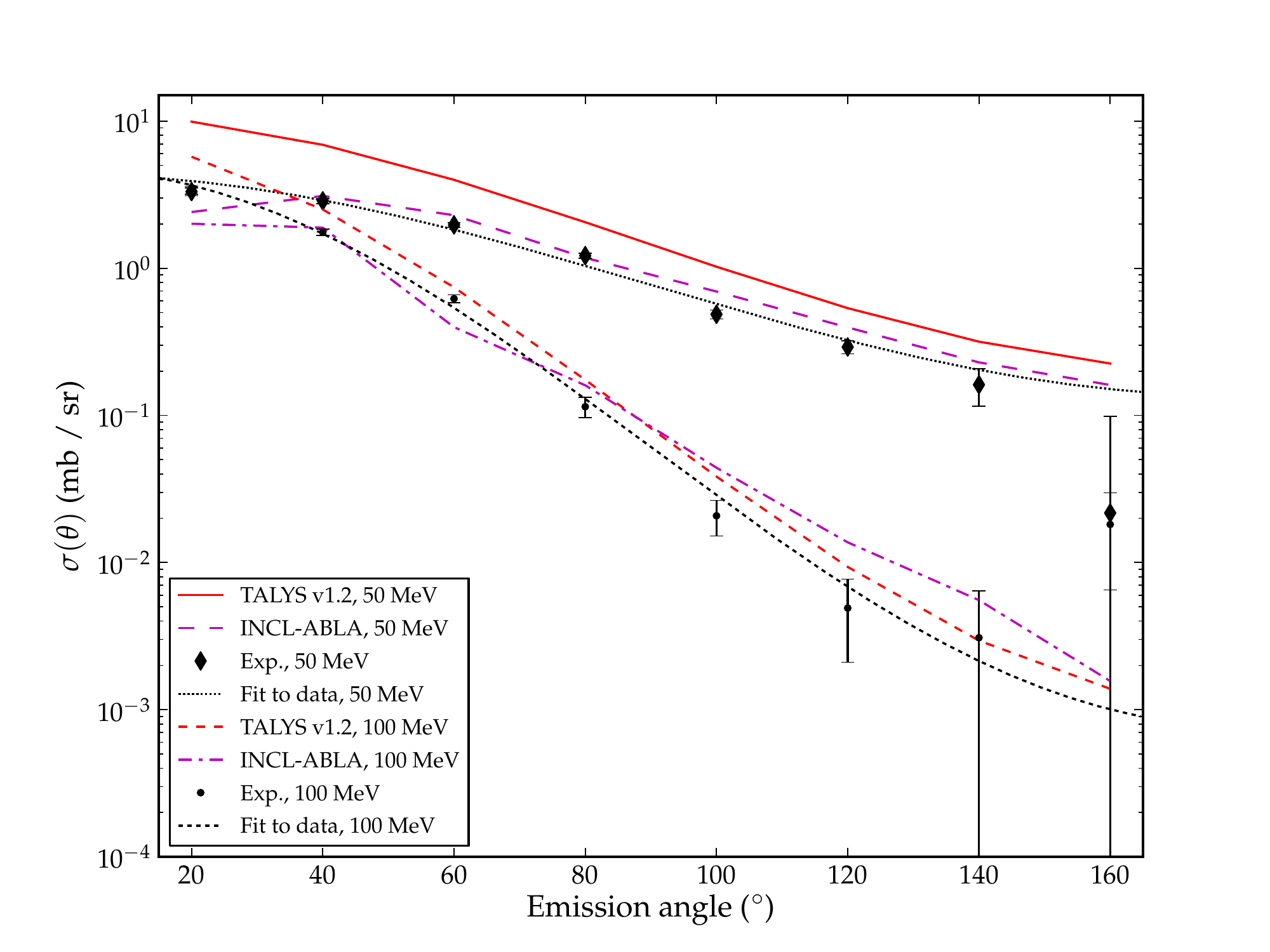}
\caption{\label{fig:bi_p_adxs2}Angular differential cross sections for the Bi(n,xp) reaction at the secondary particle energies 50\,MeV and 100\,MeV compared to TALYS and INCL-ABLA.}
\end{figure*}

The INCL-ABLA calculations are seen to agree quite well with the data although the INC model was originally developed for nucleon energies above around 200 MeV, but this limit has been extended downwards. Thanks to improved treatment of low energy projectiles, INCL4.5 gives quite good results much below 200\,MeV. The largest differences to the data are found for proton production which is underestimated at $20^\circ$. This is also visible for the angular differential spectra in Figs.~\ref{fig:fe_p_adxs1}-\ref{fig:bi_p_adxs2}.

MCNP6 which is also partly based on the INC model generally overestimates the cross sections at lower emission energies, while it instead underestimates them at higher energies. However, the $\alpha$-particle production is an exception where good agreement with the data is found for both the iron and bismuth case. The same applies to the results of the MQMD-GEM calculations which agree better with the $\alpha$-particle production than the other channels.

\subsubsection{Angular differential cross sections}
To improve visibility only the results from TALYS and INCL-ABLA calculations were plotted in Figs.~\ref{fig:fe_p_adxs1}-\ref{fig:bi_p_adxs2}. Both TALYS and INCL-ABLA agree fairly well with the angular differentiated data. TALYS still generally overestimates the cross sections while INCL-ABLA shows very good agreement with the data in these plots, except at $20^\circ$.

\subsubsection{\label{sec:edxs}Energy differential cross sections}
To obtain the energy differential cross sections angular corrections were calculated. For each particle and energy bin, a correction was obtained by fitting the data to an exponential function \cite{kal88}, $a e^{b\cos\theta}$, and extracting the extrapolated cross sections in the 0$^\circ$-10$^\circ$ and 170$^\circ$-180$^\circ$ regions. They are typically small due to the small solid angle in these regions even though the cross section is the largest close to $0^\circ$.

The INCL-ABLA calculations are predicting the cross section well for all channels for both targets (Fig.~\ref{fig:fe_edxs},~\ref{fig:bi_edxs}). At lower energies TALYS is overestimating all iron cross sections except the Fe(n,x$\alpha$) one. MCNP6 and MQMD-GEM work best for $\alpha$-production.

\subsubsection{Production cross section}
\begin{figure*}[htbp]
\includegraphics[width=0.7\textwidth]{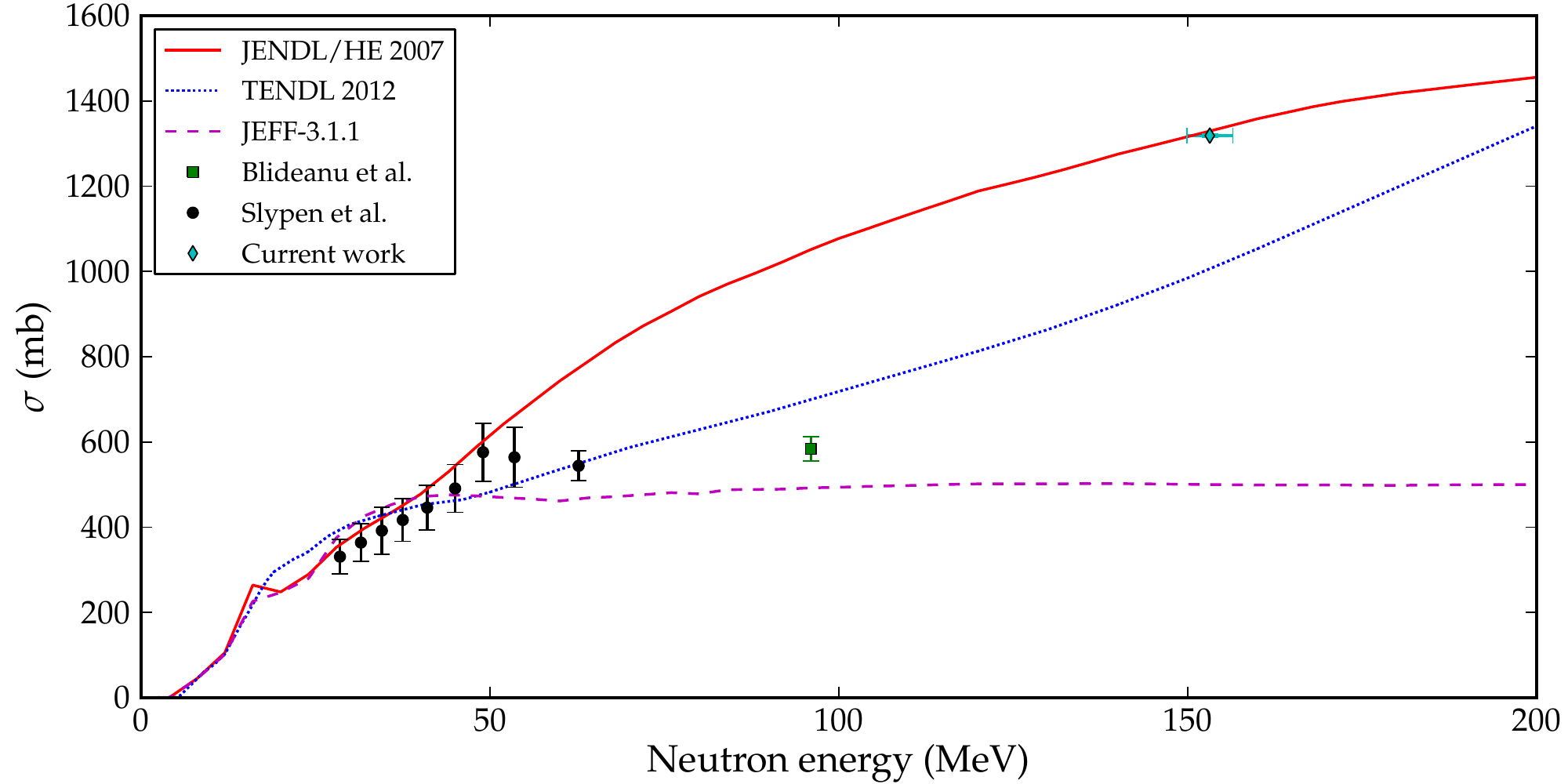}
\caption{\label{fig:fe_p_prod}Total cross section of the Fe(n,xp) reaction versus incident neutron energy. Evaluated data for $^{56}$Fe were used for the comparison even though the experimental target had natural isotopic abundance. The position of our data point on the energy axis reflects the convolution of the QMN neutron spectrum (Fig.~\ref{fig:acceptedspectrum}) and the shape of the cross section (see text for details). We compare our data to TENDL 2012 \cite{tendl}, JENDL/HE 2007\cite{jendlhe} and JEFF 3.1.1 \cite{jeff311} as well as to data from Blideanu et al.\cite{blideanu} and Slypen et al.\cite{slypen} }
\end{figure*}

In Fig.~\ref{fig:fe_p_prod} we have compared the total cross section of Fe(n,xp) to other experiments as well as the TENDL 2012 \cite{tendl} and JENDL/HE 2007 \cite{jendlhe} and JEFF 3.1.1 \cite{jeff311} evaluations. Evaluated data for $^{56}$Fe were used for the comparison even though the experimental target had natural isotopic abundance.

Our measured data point is a convolution of the neutron flux (Fig.~\ref{fig:acceptedspectrum}) and the cross section. The shapes of the evaluated cross section were used to estimate the weighted mean neutron energy in this measurement. The error bars in Fig.~\ref{fig:fe_p_prod} represent both the uncertainty due to the neutron spectrum and the uncertainty of the cross section shape. The shape uncertainty's upper and lower limits were estimated by comparing the weighted mean neutron energy obtained from each evaluation.

The previously mentioned angular correction that was calculated for the energy differential cross sections is included also for the production cross section. Our data point agrees well with the JENDL/HE evaluation, but the Blideanu et al. data point \cite{blideanu} at 96\,MeV instead favour TENDL or JEFF. Most of the data points of Slypen et al. \cite{slypen} fit all evaluations but their data point highest up in energy favours TENDL and JEFF.

\subsection{Current development of models}  
Efforts on improvement of the pre-equilibrium model used by the CEM event generator of MCNP6 were started recently \cite{mcnp-11,mcnp-12}. It is expected that after the completion of this extension of the pre-equilibrium model in CEM, MCNP6 should predict better spectra of complex particles and light fragments heavier than $^4$He emitted in various nuclear reactions, including the light-ion production by neutrons on Fe and Bi measured and discussed here. It should be noted also that in later versions of TALYS new complex-particle emission models from Kalbach \cite{kal05} have been implemented.

\section{Conclusions}
We have presented the final results from measurements of double-differential cross sections for neutron-induced light-ion production in an energy range where little data exist. The procedure for correcting for the energy loss in the targets has been validated.

The cross sections were compared to four different models which overall agreed fairly well with the measurements. Especially the $\alpha$-production is often seen to be more closely reproduced by the calculations.

The energy and angular differential cross sections were also compared to models and notably the INCL-ABLA calculations agreed very well.

The proton production cross section on Fe was compared to the TENDL 2012 evaluation which underestimated the cross section compared to the experimental data. Comparing the same production cross section to the JENDL/HE 2007 evaluation a very close agreement was found between the evaluation and the experimental data.
 
\begin{acknowledgments}
The authors would like to thank the TSL staff for their assistance during the experiments. This work was supported by the Swedish Radiation Safety Authority, the Swedish Nuclear Fuel and Waste Management Company, Ringhals AB within the NEXT Project, and by the European Commission within the Sixth Framework Programme through I3-EFNUDAT (EURATOM contract no. 036434) and within the Seventh Framework Programme through ANDES (EURATOM contract no. 249671).

Part of the work performed at Los Alamos was carried out under the auspices of the National Nuclear Security Administration of the U.S. Department of Energy at Los Alamos National Laboratory under Contract No. DE-AC52-06NA25396.

This work is supported in part (for L.M.K) by the M. Hildred Blewett Fellowship of the American Physical Society, \url{www.aps.org}.

\end{acknowledgments}

\bibliography{prc_final_draft}



\end{document}